\documentclass{article}

 \usepackage[final]{neurips_2025}

\usepackage[utf8]{inputenc} % allow utf-8 input
\usepackage[T1]{fontenc}    % use 8-bit T1 fonts
\usepackage{hyperref}       % hyperlinks
\usepackage{url}            % simple URL typesetting
\usepackage{booktabs}       % professional-quality tables
\usepackage{amsfonts}       % blackboard math symbols
\usepackage{nicefrac}       % compact symbols for 1/2, etc.
\usepackage{microtype}      % microtypography
\usepackage{xcolor}         % colors
\usepackage{enumitem}

\title{Conversational Alignment with Artificial Intelligence in Context}
\author{%
  Rachel Katharine Sterken$^{1}$ \quad James Ravi Kirkpatrick$^{2, 3}$ \\
  $^{1}$University of Hong Kong \quad $^{2}$University of Oxford \quad $^{3}$ Magdalen College, Oxford\\
  \texttt{sterkenr@hku.hk} \\
  \texttt{james.kirkpatrick@philosophy.ox.ac.uk} \\
}

\begin{document}

\maketitle

\begin{abstract}
The development of sophisticated artificial intelligence (AI) conversational agents based on large language models raises important questions about the relationship between human norms, values, and practices and AI design and performance. This article explores what it means for AI agents to be \textit{conversationally aligned} to human communicative norms and practices for handling context and common ground and proposes a new framework for evaluating developers' design choices. We begin by drawing on the philosophical and linguistic literature on conversational pragmatics to motivate a set of desiderata, which we call the CONTEXT-ALIGN framework, for conversational alignment with human communicative practices. We then suggest that current large language model (LLM) architectures, constraints, and affordances may impose fundamental limitations on achieving full conversational alignment.\\

\end{abstract}

\section{Introduction}

In recent years, advances in deep learning have driven a transformative shift in artificial intelligence (AI), leading to the development of sophisticated conversational agents based on large language models (LLMs), such as OpenAI's ChatGPT, Anthropic's Claude, Google's Gemini, Meta's Llama, and DeepSeek. 
These AI systems are capable of text-generation and other language-based tasks with unprecedented fluency and proficiency, leading some theorists to claim that sophisticated LLMs like GPT-4 have attained ``a form of \textit{general} intelligence, indeed showing \textit{sparks of artificial general intelligence}'' \citep[92]{bubeck2023sparks}. 
Indeed, LLMs have demonstrated remarkable success across a wide range of domain-specific and domain-general reasoning tasks that extend beyond mere regurgitation of training data, such as translation \citep{wang2023documentlevel}, sentiment analysis \citep{kheiri2023sentimentgpt}, question answering \citep{openai2023gpt4}, mathematical reasoning \citep{lewkowycz2022solving}, summarization \citep{wang2023documentlevel}, and code generation \citep{savelka2023thrilled}.\footnote{See \citet{milliere2024philosophical}, for a recent review.}
These systems are already being integrated into other technologies, such as search engines, customer service platforms, research and educational tools, personal assistants, AI companions, AI doctors, AI agents, and multi-agent systems, marking a fundamental change in how humans interact with AI \citep{manzini2024code, kirk2025why}.

These technological advances raise important questions about the potential risks and implications of the increasingly widespread use of conversational agents \citep{kenton2021alignment, weidinger2022taxonomy}. 
Much of the extant discussion surrounding AI alignment has centered around ensuring that LLMs adhere to human values, ethics, and safety constraints.\footnote{See \citet{gabriel2022challenge} and \citet{anwar2024foundational}, for recent overviews.}  
For example, three commonly adopted alignment criteria on which models are fine-tuned are the ``three Hs''---honesty, helpfulness, and harmlessness---which are intended to prevent them from generating misinformation, toxic language, or harmful recommendations \citep{askell2021general, glaese2022improving}. 
But although the focus on \emph{ethical alignment} is essential to mitigate misaligned AI systems that could produce biased, deceptive, or even dangerous outputs, the emphasis on moral constraints has largely overshadowed a different but equally fundamental issue of AI alignment: Are LLMs actually aligned to human \emph{conversational} norms and practices? 
Human conversation is shaped through our situated interactions with other humans in context; it requires jointly and cooperatively building and updating a shared set of beliefs, goals, and practices, the implicit understanding of the nuances of different communicative norms and contexts, and the ability to make reasonable pragmatic inferences. 
These components go far beyond simply producing grammatically correct and intelligible sentences. 
In contrast, LLMs are disembodied computational entities that acquire their capability for text-generation in the form of next-token prediction through statistical analysis of large corpuses of text. 
Even if language models are designed to be aligned with our ethical values, their constitution is fundamentally different to humans, and so they may fail to engage in conversation in pragmatically appropriate and characteristically human ways.
In turn, this may have a disruptive influence on human communicative norms \citep{kasirzadeh2023conversation}. Furthermore, ensuring that LLMs adhere to human conversational norms is not just a matter of competence or coherence, it is also a crucial aspect of safety. 
Misalignment in communication can lead to breakdowns in communication, misunderstandings, misinformation, and the erosion of trust in AI systems. 
As conversational agents become further embedded in other AI systems and deployed in high-stakes use cases like healthcare, legal advice, or crisis response, it is essential that the language models can effectively track content and context, interpret conversational nuance, and adhere to pragmatic norms.

In this paper, we examine the question of \emph{conversational alignment} in AI systems, the process of encoding human conversational values and goals into AI models to make them helpful, safe, and reliable conversational partners. More specifically, we ask: What does it mean for a conversational agent to be aligned with human communicative practices? Do LLMs follow the norms, structures, and pragmatic expectations that govern human conversation? How do LLMs differ from human speakers in their conversational abilities? Are current LLM architectures, operational constraints, and affordances fundamentally limited in achieving full conversational alignment or can they be improved to better approximate human dialogue? By shifting the focus from ethical alignment to conversational alignment, we seek to uncover the linguistic and pragmatic limitations of LLMs and their broader implications for AI--human interaction.

The plan for this article is as follows. 
Section \ref{sec:2} explores various dimensions of conversational context from the literature in philosophy of language and linguistics, connecting them to existing discussion of the disruptive potential of emerging technologies on communication. 
In Section \ref{sec:3}, we propose the CONTEXT--ALIGN framework, a general framework of fundamental conversational values grounded in semantic and pragmatic theory, to identify and evaluate conversational alignment in language models. 
Section \ref{sec:4} investigates the problem of context window overflow in large language models and argues that there is a fundamental tension between maintaining conversational coherence and avoiding context collapse. 
Section \ref{sec:5} analyzes the pragmatics of prompting. We show how static prompting are poor substitutes for a dynamically co-constructed context and critically assess how behavioral alignment protocols impose static communicative identities on LLMs that limit their pragmatic competence. 
Section \ref{sec:6} concludes by drawing out the philosophical and practical implications of our analysis and proposing directions for future research and development.

Before we begin, some preliminary framing remarks. 
In the next section, we outline central views in philosophy of language, semantics, and pragmatics with broad brushstrokes. 
The field is not characterized by consensus and convergence. 
Quite the contrary, there is a great deal of disagreement. 
There are Chomskians, who deny the existence of languages and semantics, Griceans who between themselves disagree about the details of the conversational maxims, relevance theorists who don't appeal to conversational norms, dynamic semanticists, and radical contextualists. 
Therefore, what we sketch in this paper is not an effort to capture a consensus view. 
Instead, our aim is to select the views in the field that have been central to work in this area over the last 50 years. 
Different theorists will endorse a subset of these views and emphasize different aspects. In this way, the CONTEXT-ALIGN framework proposed below could be articulated in different ways, depending on one's theoretical outlook. 
The main upshots presented here, however, would remain the same: to ensure genuine pragmatic competence and safety, and to minimize disruptive effects on human communication, developers should aim to improve conversational agents along the lines proposed in the CONTEXT-ALIGN framework.

\section{Theories of Conversational Context and Communication} \label{sec:2}

Successful communication requires that conversational participants share a common set of background assumptions and interpretative tools that are needed to understand one another's conversational contributions as part of a cooperative endeavor. 
In this section, we introduce several ways of conceiving of conversational context and its import to communication, drawing on resources from philosophy of language and linguistics. 
We shall focus on the context-sensitivity of semantic content, pragmatics, common ground, conversational score, information structure and Questions Under Discussion (QUDs), and context collapse. 
These notions will then inform our understanding of what it is for conversational agents to be conversationally aligned with the demands of context on successful communication.

\subsection{Semantic Content and Context-Sensitivity}

The standard view in philosophy of language and linguistics is that a theory of meaning should assign the content (or semantic value) of a complex expression or sentence of a language as a function of the assignment of content (or semantic values) of its constituents and its syntactic structure. 
Furthermore, it is widely agreed that the content of an utterance of a sentence is determined by the conventional meaning of the expressions in that sentence relative to the context in which the sentence is uttered \citep{montague1968pragmatics, montague1970english, lewis1970general, kaplan1989demonstratives}. 
For example, if Rachel utters the sentence \textit{I am here now}, she expresses that she is in Hong Kong at 12:00 on 12th February 2025, but if James utters that sentence, he expresses that he is in Oxford at 20:00 on 12th February 2025. 
Here, the semantic contribution of the indexical expressions like \textit{I}, \textit{here}, and \textit{now} to the content of these utterances are determined by the speaker, the location, and the time of the utterance of the sentence in which those expressions occur. 
Context-sensitivity arguably extends beyond indexicals to other expressions, such as gradable adjectives \citep{kennedy2007vagueness}, modals \citep{kratzer1977what}, conditionals \citep{stalnaker1968theory, lewis1973counterfactuals}, quantifiers \citep{stanley2000quantifier}, generics \citep{sterken2015generics}, and vague expressions \citep{graff2000shifting}. 
For example, the content of an utterance of a sentence like \textit{John is tall} may depend on the contextually salient threshold on the height scale \citep{kennedy2007vagueness}. 
In the context of five-year olds, an utterance of the sentence \textit{John is tall} may express that John is tall for a five-year old, while in the context of NBA basketball players, it may express that John is tall for an NBA player. 
Consequently, successful communication requires conversational participants to be able to track the relevant features of the context for interpreting context-sensitive expressions. 

\subsection{Pragmatics}

There are several ways in which speakers imply things or hearers infer things that go beyond the literal, conventional meanings of the sentences that the speaker utters. 
For example, if a parent asks their child whether they have finished their dinner and the child replies \textit{I have eaten my carrots}, the parent might infer that the child still has food left to eat, even though that was not part of the literal meaning of the child's utterance. 
Or the child might convey that they disliked the dinner by a sarcastic remark like \textit{That was sooo tasty\ldots}. 
Following the influential work of \citet{grice1975logic}, many theories of pragmatic meaning have been developed to study and explain how such meanings are communicated and how this is shaped by speaker intentions, conversational norms, and hearer inferences (see, e.g., \cite{horn1984new, sperber1986relevance, recanati1989pragmatics, levinson2000presumptive, lepore2015imagination}). 
A central theme running through all of these approaches is the assumption that speakers are tacitly understood by hearers to be following general pragmatic principles or conversational norms that govern cooperative conversation, and so they draw inferences on this basis by noting when speakers are being cooperative and when they are not. 
Consequently, successful communication requires hearers to be attuned to whether speakers are being cooperative and to be able to draw the appropriate inferences accordingly.

In addition to making such implications, speakers also use language to perform \emph{speech acts}, such as making requests, issuing commands, or giving promises \citep{austin1962how, searle1969speech}. 
In some cases, saying something is the case is enough to make it so. 
For example, we make you a promise by uttering the words \textit{We promise you that we'll do such and such}. 
But speech acts often require that certain felicity conditions are met for the speech act to be successful. 
For example, you cannot pronounce a couple married if you lack the authority to do so, nor can you bet us on the outcome of a contest unless we accept your bet. 
Furthermore, conversational participants must be fluent in the constitutive norms governing speech acts. 
These norms partly determine if and when an attempt to make a particular speech act has been successful. 
Consequently, successful conversation often requires that conversational participants are in a position to make and can recognize speech act attempts.\footnote{For an extensive discussion of pragmatic norms and their relation to the design of conversational agents, see \citet{kasirzadeh2023conversation}. See also work on evaluating the basic pragmatic capabilities of LLMs \citep{sravanthi2024pub, ma2025pragmatics}.} 

\subsection{Common Ground}

Keeping track of contextual information is necessary for the proper interpretation of context-sensitive expressions, what is said and communicated, and to help with pragmatic reasoning. 
Following \citet{stalnaker1968theory, stalnaker1978assertion, stalnaker1999context, stalnaker2002common, stalnaker2014context}, many theories of conversational context are taken to include a \emph{common ground}, a background of information that is presumed to be shared by the conversational participants and against which the interpretation of conversational contributions takes place. 
Suppose that we agree that it will be sunny today, but we have not yet agreed whether it will be warm enough to go to the beach. 
Then it is common ground between us that it will be sunny today, but our mutually presupposed beliefs won't settle the question about whether it is warm enough to go to the beach. 
The common ground also dynamically evolves over the course of a conversation. 
On the Stalnakerian picture, an assertion functions as a proposal to update the shared assumptions between conversational participants with new information. 
So as new assertions are made and accepted, the common ground is updated to align with the informational content of those utterances. 
In turn, the updated common ground serves as background for the interpretation of new conversational moves. 
So if we now say that it is warm enough to go to the beach and you agree, then our common ground will be updated to include the information that it is warm enough to go to the beach.

Keeping track of common ground is crucial for successful communication. Suppose we are trying to decide where to go for dinner this evening. An utterance of the sentence \textit{We should go there} will not make much sense unless it is settled what \textit{there} refers to and this referent is available in the common ground. That is, if the utterance is discourse initial and there is no plausible candidate for the referent of \textit{there} in the common ground, the conversation crashes. Equally, successful communication changes the common ground. For example, if I say \textit{Lao Ma Spicy is tasty}, it becomes common ground between us not only that I uttered the sentence \textit{Lao Ma Spicy is tasty}, but also that there is a suggestion to go there to eat, with Lao Ma Spicy becoming a plausible referent of \textit{there}. Consequently, if you say \textit{We should go there}, the referent of there is now settled by the common ground. 

Similarly, failure to make appropriate adjustments to one's beliefs as conversation proceeds can lead to misrepresentations of the common ground between conversational participants. For example, if John believes that the murderer of Smith is Jones, while Mary believes that the murderer of Smith is Brown, then it won't be common ground who the murderer of Smith is. In turn, this can lead to misunderstandings as conversation that uses the description \textit{the murderer of Smith} progresses, since John will be updating his beliefs about Jones, while Mary will be updating her beliefs about Brown. These misunderstandings can only be rectified by resolving the disagreement about who the murderer of Smith is. In summary, successful conversation requires an ability to keep track of the common ground.

\subsection{Conversational Scoreboard}

Communication serves a wider range of purposes beyond mere inquiry. Some communication is non-propositional. For example, saying \textit{I'm so excited} or \textit{This is frustrating} expresses emotional states, in addition to propositional information. 
Similarly, speech acts like exclamations and interjections (e.g., \textit{Wow!}, \textit{Ugh!}) also function in this way. Communication is also used to shape attitudes other than beliefs. For example, commands and requests aim to make some act in a certain way rather than simply informing them about some information. Language is also crucial for forming interpersonal connections and group identity. The compulsion of British people to talk about the weather is better understood as phatic communication aimed at establishing and building social relations, rather than symptomatic of perineal meteorological ignorance. Communication also enables the coordination of cooperative and joint action. We use language to organize, plan activities, negotiate, and give instructions. Consequently, our representations of communicative context must be rich enough to keep track of the various uses of communication.

Following \citet{lewis1979scorekeeping}, theories of conversational contexts are taken to build in the notion of a \emph{conversational scoreboard}. Consider the scoreboard in a baseball game: it tracks categories of information, such as runs, strikes, and outs, thereby determining the current state of play and the set of permissible future moves. Similarly, language can be considered as a game \citep{wittgenstein1953philosophical}: think of communication as a conventional, rule-governed practice, where the kinds of intelligible or permissible actions are determined by the constitutive rules of the practice. According to this analogy, the conversational scoreboard keeps track of the conversational moves or speech acts that are performed or attempted by the conversational participants, as well as other relevant features such as discourse referents, standards of precision, facts about salience, or spheres of permissibility. The scoreboard also determines which conversational moves count as felicitous, although Lewis points out that the scoreboard can be updated to make an infelicitous move count as felicitous through a process called \emph{accommodation}. 

To see how the conversational scoreboard is updated and revised as the conversation progresses, consider the following conversation:

\begin{enumerate}[label=(\arabic*)]
\item \label{ex:1} Alice: I saw John at the library today.\\
Bob: Studying won't help him with his exam tomorrow.\\
Alice: Oh, he wasn't studying. He was reading a novel. 
\end{enumerate}
In this conversation, we can see how each exchange updates the scoreboard to adjust the informational content and mutually held assumptions to maintain coherence in the conversation. First, the scoreboard is updated to include the fact that Alice saw John at the library, thereby introducing John as a discourse referent for future anaphora. When Bob replies, the scoreboard is updated with a presupposition that John was at the library for academic purposes. Finally, this presupposition is revisited, and the scoreboard is updated to reflect that John was reading for leisure. In general humans naturally repair misunderstandings in conversation in different ways, such as through clarifications like \ref{ex:1}, as well as self-corrections (e.g., \textit{Oops, I meant next Friday, not this one}) and requests for elaboration (e.g., \textit{Can you explain what you mean?}). These mechanisms of conversational repair can also be modelled using the notion of common ground and conversational score. In summary, the constitutive, rule-governed, and dynamic nature of communication requires the ability to model and update the conversational score.

\subsection{Information structure and QUDs}

Human conversation is not merely a sequence of disconnected statements, but rather goal-oriented joint ventures structured around what formal pragmatists call \emph{Questions Under Discussions} (QUDs). Building on the work of \citet{lewis1969convention, lewis1979scorekeeping} and \citet{stalnaker1978assertion}, \citet{roberts2012information} argues that conversations are organized around implicit and explicit questions that guide the flow of discourse. These questions help to establish shared conversational goals to ensure that participants contribute relevant and meaningful information. For example, if we are discussing where to go for dinner, our conversation will naturally revolve around the overarching question, \textit{Where should we eat?}. This QUD partly determines what counts as ``correct play'' in our conversation. For example, if you say \textit{Lao Ma Spicy is nearby}, you will be understood as proposing Lao Ma Spicy as a place to eat. QUDs also play a role in topic coherence, as they provide a mechanism for tracking what information is relevant to the conversation at hand. For example, if you say \textit{My knee is hurting}, you will be understood as making a request that our culinary destination shouldn't be too far away. But if you say \textit{Route 20 is the longest road in the US}, your contribution will not be understood as relevant to the point of the conversation. More recently, theorists have argued that the QUD not only structures the discourse, it also partly determines \emph{what is said} by an utterance of a sentence \citep{schoubye2016what}. Without the guiding structure of QUDs, conversations would risk becoming disjointed and inefficient, with participants failing to keep track of the structure of discourse, as well as failing to recognize the relevance and even the semantic content of each other's conversational contributions.

\subsection{Context Collapse}

Successful communication requires conversational participants to keep track of numerous features of the context. But advances in technology have significantly changed how we communicate with one another and they have the potential to disrupt our ability to track important features of conversational environments. For present purposes, we shall focus on one specific way that technology has impacted our communication, namely, \emph{context collapse}. Context collapse is the flattening and merging of multiple distinct audiences, each with their own accompanying social norms and expectations, into a single communicative context, which in turn makes it difficult for speakers to tailor their message appropriately \citep{meyrowitz1985no, marwick2011tweet}. The concept was developed from the ``situationist'' theory of Erving Goffman, according to which changes in human behavior are a factor of situational contexts rather than internal traits of the individual. Goffman observed that we often present different versions of ourselves and vary our form of expression across different situations and between different audiences, in part to reflect variation in our audience's identities and the social norms that apply, but also to accentuate and minimize different characteristics of ourselves \citep{goffman1959presentation}. 

In offline contexts, this need not result in any tensions or perceived inauthenticity, since different situations can be kept separate. But technological advances brought about by electronic media like radio, television, and the Internet complicate this picture \citep{meyrowitz1985no, marwick2011tweet}. In \emph{No Sense of Place} (1985), Joshua Meyrowitz, building on Goffman's work, notices that, while offline communication occurs in physically bounded contexts limited only by the barriers of perception (i.e., who can be seen and heard by whom), electronic media radically undermines and disrupts the ``traditional relationship between physical setting and social situation'' \citep[7]{meyrowitz1985no}. In turn, this disruption leads to a fusion of previously bounded communicative contexts, social norms, and ``potentially infinite degrees and patterns of situational overlap'' \citep[42]{meyrowitz1985no}. Consider the example of Stokely Carmichael, the civil rights activist involved in the Black Power movements. When engaging with Black and White audiences separately, Carmichael would tailor his rhetoric to reflect the varying sympathies between these two audiences to avoid alienating them from his causes. But when his speeches began being broadcast on TV and radio, these separate audiences were flattened into a single context and Carmichael could no longer tailor his rhetoric in different ways. Ultimately, he decided to stick with his rhetoric for primarily Black audiences.

Drawing on Meyrowitz's work, dana boyd and her collaborators extend the concept of context collapse to social media platforms like Facebook and X (formerly known as Twitter) \citep{boyd2002faceted, boyd2010social, boyd2013how, boyd2014its, marwick2011tweet}. Just as TV and radio forced Carmichael to contend with previously distinct audiences in a single communicative context, social media also forces us to ``contend with groups of people who reflect different social contexts and have different expectations as to what's appropriate'' \citep[50]{boyd2010social}. Additionally, the persistence, replicability, and searchability of information on social media platforms mean that we can never be certain about who the potential future audiences of our posts will be. Marwick and boyd argue that context collapse on social media can give rise to crises of identity and authenticity, since we cannot simultaneously satisfy the various norms governing our interactions with friends, family, and colleagues when all of these groups are potential audience members online.

Context collapse also has important ramifications on how we communicate. For example, \citet{frost-arnold2021epistemic, frost-arnold2023who} argues that, in cases of context collapse, unintended viewers of a post may fail to share information about the poster and the norms and goals of the conversation, which means that they are unlikely ``to recover the proposition that the speaker intended to convey to their imagined audience'' \citep[444]{frost-arnold2021epistemic}. Furthermore, Karen Lewis \citet{lewis2025imagined} argues that online conversations and context collapse present a challenge for the centrality of common ground in theorizing about conversational contexts, both in terms of its static role as the background against which interpretation takes place, and its dynamic role as the thing updated by conversational moves:

\begin{quote}
Collapsed contexts don't just collapse social contexts, but also potential common grounds for a conversation. \ldots Speakers on social media also have to navigate what, if anything, they can take for granted as background to a post, given the constant potential for collapsed contexts\ldots.

Online environments do not lend themselves to making any particular speech act ``obviously evident to all''. \ldots [F]or many who post online, there is nothing that guarantees that the fact that a post was made is obviously evident to the relevant parties\ldots Posting online is much more like making an assertion by shouting into a loud, crowded auditorium, not knowing if anyone will hear, or who it will be if someone does \citep[6--8]{lewis2025imagined}
\end{quote}

Similarly, Lucy McDonald \cite{mcdonald2025context} describes context collapse in online communication as resulting from multiple overlapping language games, which she likens to trying to play a baseball game in a park that is busy with many other ball games going on and where moves are happening all over the place:

\begin{quote}
Online communication often involves multiple overlapping language games occurring simultaneously, with indeterminate contours and an ever changing set of participants, only ever partially known to the other players. Different contexts collapse into one another not only in the sense that different audiences come together, but also in the sense that there are multiple scoreboards and multiple distinct sets of evolving norms. \citep[18]{mcdonald2025context}
\end{quote}

In summary, context collapse can have the undesired effect of generating conversational breakdowns or miscommunication, say when conversational participants fail to know enough about the communicative context to adequately model the conversation they are in or to determine what is being communicated.

\subsection{Interim Summary}

The preceding sections outline five foundational elements of conversational context---context-sensitivity, pragmatics, common ground, the conversational scoreboard, and information structure and QUDs---and demonstrate how these components are both theoretically structured and practically challenged by evolving communication technologies. More specifically, we have shown how successful communication relies on our ability to track contextually salient features of the context, changes in the common ground and conversational score, and the mutual goals of the conversation. Furthermore, we have seen how technologies that mediate communication can disrupt these abilities. In particular, in the case of context collapse on social media, important features of the conversational context can be obfuscated, and the stability of shared audience expectations may be fractured. In turn, this complicates the establishment of common ground and a shared conversational score.

With the rise of LLMs as conversational agents, new questions emerge about how their architectures, operational constraints, and affordances reshape communicative contexts. Unlike human interlocutors, LLMs lack appropriate contextual situatedness, genuine intentionality, and the rich socio-linguistic competence that informs human pragmatic inference. Their ``understanding" of context is mediated through statistical patterns in training data and the technical parameters of their design, such as finite context windows, token limits, and alignment protocols. In turn, this raises concerns about whether LLMs can truly participate in the mutual construction of shared context, whether they even track such features, and whether their outputs distort conversational dynamics.

\section{The CONTEXT-ALIGN Framework} \label{sec:3}

To assess whether LLMs can achieve conversational alignment with human practices, we propose the CONTEXT-ALIGN framework, a set of desiderata grounded in the theories of context surveyed in the previous section.\footnote{  Dafoe et al. \cite[19--20]{dafoe2020open} list communication and common ground as open problems in cooperative AI. Kasirzadeh and Gabriel \cite[6]{kasirzadeh2023conversation} discuss context as important for conversational alignment.} These criteria aim to ensure that AI systems function as competent, reliable interlocutors, while minimizing disruptive effects on human communicative norms. The CONTEXT-ALIGN framework differs from important benchmarks that have recently been developed to test the basic pragmatic capabilities of LLMs, such as implicature recovery and figurative language understanding \citep{sravanthi2024pub, ma2025pragmatics}. Such benchmarks are a crucial first step towards conversational alignment. The CONTEXT-ALIGN framework seeks to build on such benchmarks (see CA6 below) and the more normative work of Kasirzadeh and Gabriel \cite{kasirzadeh2023conversation} on conversational alignment to secure what we see as other important factors in securing conversational alignment.

\begin{enumerate}
    \item \emph{Tracking of semantic content and context-sensitivity}: The model must dynamically interpret context-sensitive expressions (e.g., indexicals, anaphora, gradable adjectives) relative to the evolving conversational context, including user identity, spatiotemporal parameters, and conversational goals.
    \item \emph{Common ground management}: The model must be able to adequately manage the common ground by recognizing how conversational contributions change mutually shared assumptions, as well as keeping separate the common ground of different conversations.
    \item \emph{Conversational scoreboard updating}: The model must be able to dynamically update the conversational score with each conversational contribution, effectively accommodate violations of `correct' or felicitous play, and repair conversational scoreboards in the face of communicative breakdowns.\footnote{  CA2 and CA3 presuppose the ability to identify and classify speech acts as they are performed by the user, as well as the speech acts the LLM itself is performing. A further set of benchmarks is needed to evaluate these capabilities.} 
    \item \emph{Discourse and information structure management}: The model should be able to adequately manage the dynamics of QUDs, topic and discourse structure, relevance, and salience. 
    \item \emph{Capacity for accommodation}: The model should accommodate repairs and clarifications seamlessly, updating its understanding of context without requiring restatements of prior information.
	\item \emph{Pragmatic inference}: The model should be able to generate and interpret implicatures, speech acts, and other non-literal language consistent with human norms. 
    \item \emph{Ethical--pragmatic integration}: Alignment with ethical principles (e.g., honesty, compassion) must not come at disproportionate cost to the pragmatics of conversation, nor to successful communication. 
    \item \emph{Context-collapse and context-distortion mitigation}: The model should recognize and adapt to collapsed or distorted contexts by inferring or querying the user's intended audience, register, and norms. This includes avoiding conflating incompatible linguistic practices.
    \item \emph{Identification of defective contexts and conversational structures, and repair and clarification protocols}: The model should proactively signal misunderstandings or breakdowns in context and conversational structure, and solicit user input to resolve ambiguities.
    \item \emph{Transparency in context-handling}: Users should be able to query the model's understanding of the current context and conversational properties to correct misalignments.
    \item \emph{Cross-contextual memory}: The model should retain and apply information from prior interactions (within privacy constraints) to establish continuity and depth in recurring conversations.\footnote{Each of the criteria in the CONTEXT-ALIGN framework presupposes other important capabilities. They require the ability to implement a theory of mind, that is, the ability to identify, classify and reason about users' mental states (see, a.o., \cite{aru2023mind, sap2023neural, street2024llm, street2024llms, ullman2023large, wang2024mutual, williams2022supporting}). It also needs to use that to adjust its own states, so there will be a need for extensive higher-order reasoning. This includes, importantly in the case of communication, a suite of metalinguistic capabilities, and metacommunicative/metapragmatic capabilities.  } 
\end{enumerate}
In the remainder of this paper, we shall suggest that there are fundamental limitations of the abilities of LLMs to appropriately handle the desiderata of the CONTEXT-ALIGN framework.\footnote{Note that the desiderata in CONTEXT-ALIGN is not intended to be exhaustive of the ways that context can impact conversational alignment.}

\section{Language models and context window overflow} \label{sec:4}

In this section, we argue that there is a fundamental tension between the mechanisms required for conversational agents to sustain coherent conversations and avoiding context collapse.

\subsection{Language Models: An Overview}

Let us begin with a brief overview of a core component of contemporary conversational agents: large language models. LLMs, such as GPT-4 \citep{openai2023gpt4}, Claude 3 \citep{anthropic2024claude}, and Gemini Ultra \citep{team2024gemini}, are AI systems designed to generate and interpret human-like text. 
They operate using deep learning architectures, primarily based on transformers, which allow them to process vast amounts of textual data and predict the most likely next word in a given sequence. 
LLMs are designed and trained to recognize patterns in extensive datasets, including books, articles, and online content, using techniques like self-supervised learning and reinforcement learning from human feedback (RLHF) to refine their outputs. 
This allows them to generate coherent and contextually appropriate responses to diverse prompts or questions. The model is then fine-tuned to be more effective at following instructions and to take feedback from human evaluators to reduce toxicity, bias, and other problematic features. The resulting LLM can then be embedded in a conversational agent that can engage in conversation with the user.

Let's break down how LLMs process text and generate outputs. Transformer architecture is a deep learning model introduced by \citet{vaswani2017attention} that enables LLMs to process and generate text efficiently. Unlike earlier models that processed text sequentially and struggled with long-range dependencies, transformers use self-attention mechanisms to weigh the importance and relevance of different words in a sentence, regardless of their position or distance from other words. These mechanisms allow the model to construct sophisticated representations of long sequences of texts, by considering the relationships between all words in a sequence and the broader context in which expressions occur. Transformers are particularly powerful due to their multi-layered attention system. Each layer in a transformer applies self-attention followed by feedforward processing, allowing the model to refine its representation of the text at different layers of abstraction. Lower levels capture basic syntactic patterns, like word order and grammar, while higher layers extract semantic meaning and pragmatic context. This allows models to process complex passages of text and generate meaningful responses. 

The smallest unit of language that a model can process is called a \emph{token}. 
While tokens can map onto whole words, the correlation between words and tokens is complex, and tokens are sometimes mapped onto smaller parts of words. 
Before a sequence of text can be processed, it must first be converted into tokens through a process called tokenization. 
For example, a single word (e.g., \textit{cat}) may be tokenized as a single token, while other words may be broken into multiple tokens (e.g., \textit{unhappy} may be tokenized as \textit{un-} and \textit{happy}). 
Additionally, spaces, newline characters, and punctuation may also be included in tokens. 
Tokenization varies from language to language and from model to model. 
Every model has an upper limit on the number of tokens in the input prompt plus the number of tokens in the generated output of the model that it can process. 
This limit is called the \emph{token limit} or the \emph{context window limit}. 
Intuitively speaking, the context window is the portion of the conversation that the model can ``see''. 
As new tokens are added (e.g., from each turn in the conversation), the window shifts and older tokens pass outside the window. 
A single exchange between the user and the model consists of an input prompt and the generated output of the model. 
This is called a \emph{turn}. 
For example, a turn might consist of a user's input prompt \textit{What is the capital of France?} and the model's generated output \textit{Paris}. 
A \emph{conversation} is a sequence of turns. 
For example, the user may continue the conversation by asking \textit{Is it big?} to which the model may reply Yes, it has \textit{2.1 million people}. 
Importantly, unlike human-to-human conversation, language models process multi-turn conversations by taking, for any given turn, \emph{both} the user's specific prompt at that turn \emph{and} the inputs and outputs of all previous turns within the token limit as input.

\subsection{Context Window Overflow}

Now let us introduce the problem of context window overflow. Context window overflow occurs when the conversation exceeds the context window limit. When this happens, older tokens are pushed out of the window or sometimes the output is truncated. This has serious implications for conversational alignment as the model may lose track of prior context and important details from earlier conversation might be ``forgotten'', which in turn could lead to incoherent or confused responses, as well as a destabilization in the common ground. 

To illustrate, consider anaphora resolution. Anaphora is the use of a word to refer back to an entity previously introduced in conversation or text. Consider the following sentence:

\begin{enumerate}[label=(\arabic*)]
\setcounter{enumi}{1}
    \item \label{ex:2} John is going to the store. He is smoking.
\end{enumerate}
In \ref{ex:2}, the pronoun \textit{he} in the second sentence anaphorically refers to John. When context window overflow occurs, the number of tokens in the conversation exceeds the context window limit. When this happens, the model loses access to earlier text in the conversation, which can interfere with anaphora resolution. For example, if the model can no longer access earlier conversation, it might not ``know'' what \textit{he} refers to. Failures of anaphora resolution can lead to incoherent responses.

Another area in which context window overflow can also lead to a breakdown is topic coherence. Effective communication requires that speakers stay on topic by maintaining focus on what questions are under discussion. Humans naturally track the topic of discussion over time, using implicit cues and shared background knowledge to connect ideas and sustain thematic continuity. However, when a language model reaches its context window limit, it may lose access to earlier parts of the conversation, making it prone to drift off-topic or to produce sudden, seemingly irrelevant responses. For example, if a user is discussing climate change and asks a follow-up question about carbon emissions, the model should recognize this continuity in the topic. But if earlier context has been pushed out of the context window, the model may fail to link the new query to the established discussion, generating an off-topic or generic response. This not only disrupts the conversational flow, but it also diminishes the user's trust in the models' ability to sustain meaningful dialogue.

A closely related issue is diachronic consistency, that is, a speaker's ability to remain and be perceived to remain internally consistent across an extended interaction. In human-to-human conversation, speakers retain memory of their past statements and adjust their responses accordingly, ensuring that their own contributions do not contradict or undermine previous assertions, or when there is need to revise one's views, doing so explicitly so that the conversational common ground can be jointly revised. However, unlike humans, language models lack persistent memory that goes beyond the context. This means that once older parts of the conversation have passed outside the context window, the model cannot reliably maintain consistency with those statements. This can result in contradictions, redundant explanations, or misalignment with prior commitments. For example, if an LLM is helping a user to troubleshoot a technical issue, it may initially recommend one solution, but once that solution has left the context window, it might unknowingly suggest a second conflicting approach. Another example involves applications like AI companionship or long-form discussions, where a model might lose track of established user preferences because past interactions have fallen outside the context window. Like with failures of topic coherence, issues with diachronic consistency may lead to frustrating and disjointed exchanges, as well as concerns about the trustworthiness and usability of LLMs in extended use cases.

There are various strategies that can be used to mitigate the challenges of context window overflow, each with their own limitations. First, \emph{context compression} can be used to condense earlier parts of the conversation into a short summary that is repeated at the end of outputs \citep{askell2021general, wingate2022prompt, snell2022learning, chevalier2023adapting, wang2024incontext}. For example, at the end of each turn, the model may generate a summary that captures the key information from the conversation so far, such as the QUD, user preferences, and important named entities. This ensures that essential information can be retained and ready-to-hand without filling the context window by regularly repeating the information. Context compression typically involves keyword extraction to ensure that important terms are preserved, as well as the prioritization of core details over minor or redundant parts of the information. However, although context compression is a useful strategy, it has some limitations. Summarizing complex information often introduces loss, as some subtle contextual clues, nuanced details, or implicit relationships between pieces of information may be omitted in the process. Additionally, models may exhibit compression bias, whereby the information that the model prioritizes does not always align with conversational or user priorities. Consequently, context compression is best suited for short exchanges or structure use cases, but it is less reliable for long, multi-turn conversations involving evolving topics. As a result, context compression should not be used in isolation but rather used in tandem with other strategies for handling context window overflow.

Another strategy for handling context window overflow is to increase context window size, allowing the model to retain more conversation history within a single processing sequence. A larger context window size means that more past information and context can be retained for reference. However, this strategy also has its drawbacks. Expanding context window size also increases computational costs, as a large sequence of tokens require greater processing power and memory. These costs are often passed on to users, as most closed source LLMs operate on a pay-per-token pricing model, making this solution less practical. Furthermore, increasing context window size does not solve the problem of prioritization. Even with more text available, models still need additional mechanisms to emphasize the most relevant parts of long conversations.

Finally, external memory and retrieval systems can be used to manage context overflow by storing and retrieving information beyond the limits of the immediate conversation window. One approach is to use long-term or session-based external memory storage, which enables models to persistently record key details from past interactions and retrieve them as needed when needed. Another approach is Retrieval-Augmented Generation (RAG), which enables models to access external knowledge bases, database, or document repositories to retrieve relevant information dynamically \cite{lewis2021retrievalaugmented}. With either method, a model can recall and retrieve relevant features of the context, such as user preferences, past conversation snippets, or other stored facts, without relying only on information within the current context window. The effectiveness of retrieval-based systems depends on the quality of the retrieval mechanism.

While strategies like context compression, large context window size, external memory, and retrieval systems help mitigate context window overflow, they introduce a new challenge: context collapse. LLMs do not maintain a structured, human-like model of the conversation but instead manage context as a single long string of tokens from previous turns. As the conversation progresses and information accumulates, the model may struggle to distinguish between distinct conversational threads, to separate communicative intentions, or to maintain adequate division between different aspects or personas that a single user might adopt. For example, in a technical support chat, if a user first asks about a software bug and then later asks about account billing, an LLM that retains too much information may blend these contexts, mistakenly responding to the billing question with troubleshooting steps meant for the software issue. Similarly, in a long personal conversation with an AI conversational companion, if a user initially discusses their interest in philosophy and later shifts to talking about their vacation, an overly context-retentive LLM might inappropriately insert references to philosophical ideas in the middle of the travel discussion. Humans easily recognize these conversational shifts for what they are, namely, a user flitting between topics. But a conversational agent may not recognize that these different tasks need to be compartmentalized, which can lead to mixing up references, collapsing different subtopics, or generating responses that blur separate contexts together. The resulting phenomena is similar to context collapse in social media, where different aspects of a user get collapsed into a single interpretive frame. At the same time, if an LLM remembers too little, it faces a different set of problems: it may lose track of the common ground, fail to appropriately update on or accommodate new information, or struggle to maintain continuity with the QUDs. A model that forgets too much risks producing repetitive, inconsistent, or pragmatically inappropriate responses, failing to engage in conversation as humans do. 

Consequently, we face the following dilemma: if a conversational agent remembers too much context, it could lead to context collapse, but if it remembers too little context, then it may lose coherence, forgetting important details, which may lead to a fragmented and incoherent dialogue. Navigating these competing pressures is crucial for conversational alignment: the need to balance memory retention with context structuring to ensure meaningful, contextually appropriate, and conversationally aligned interactions. Resolving this challenge is essential for ensuring that AI systems function as reliable and pragmatically capable conversational agents.

\subsection{Summary}

To summarize, we suggest that the challenges concerning context window overflow and context collapse illustrate broader concerns for ensuring the conversational alignment of LLMs. As outlined in the CONTEXT-ALIGN framework, LLMs must track certain features of the conversational context, such as anaphoric relations (CA1), topic and discourse structure (CA4), and other features of the common ground (CA2). But context window overflow can cause models to ``forget'' prior context, leading to failures in anaphora resolution, topic coherence, and QUD tracking. Attempts to mitigate against these issues by integrating enlarging context windows or conversational memory introduces a new risk: context collapse (CA8). This trade-off exposes a fundamental tension in building LLMs for conversational alignment. Consequently, addressing these challenges requires a balance between memory retention, structured context management, and adaptive conversational repair mechanisms to prevent LLMs from either losing coherence or collapsing distinct conversational contexts.

\section{Pragmatics of Prompting and Conversational Alignment} \label{sec:5}

In Section \ref{subsec:5.1}, we suggest that prompting is a distinct speech act that serves as a novel yet problematic substitute for the dynamic, jointly-constructed context of human communication. 
In Section \ref{subsec:5.2}, we argue that, in contrast to human speakers, behavioral alignment strategies impose a rigid, pre-defined communicative identity on LLMs, which results in context collapse.

\subsection{Prompt as Utterance and Context} \label{subsec:5.1}

In the context of conversational agents, the prompt is the means by which users make their conversational contributions. Prompts stand in stark contrast to the utterances or speech acts made in human-to-human conversation. The prompt is a novel kind of communicative contribution that often serves as both an utterance and as a substitute for shared context. In human conversation, context is partly given (i.e., part of the common ground) and partly co-constructed incrementally through a series of mutually accepted conversational moves and contributions. With LLMs, however, users must scaffold context unilaterally through explicit prompting, which essentially consists of explicitly describing features of the context and what the user deems relevant to the conversation they hope to have with the LLM. In this way, prompting often simultaneously provides approximate descriptions of shared assumptions for the common ground, assignments of social-communicative roles to the user and the conversational agent, and an indication of the purpose of the conversation. The prompt serves, then, as both utterance and context---used as a more-or-less static initial frame for simulating the dynamic context and common ground that human interlocutors would otherwise have at their disposal to negotiate in-situ and iteratively.

Users instinctively engage in \emph{prompt engineering} to compensate for the LLM's inability to infer implicit contextual parameters and common ground. Prompt engineering serves to scaffold the context to increase the likelihood that the LLM's output will meet the user's expectations for the exchange. For example, a user might begin a conversation with:

\begin{quote}
\textit{Assume you are a historian specializing in 18th-century trade routes. Explain the significance of the Silk Road in this context, using non-technical language for a high school student.}
\end{quote}
Here, the user attempts to establish the model's role, the topic's scope, the intended audience, and the register all within a single prompt. In a certain sense, this mirrors the way that humans prime conversations (e.g., \textit{Let's talk about the trip we planned}), but there are crucial differences. Firstly, in human conversation, priming is largely backgrounded, implicit, or inferred directly from the context, whereas in LLM prompting, such elements need to be articulated explicitly, often in what might be understood as a null or indeterminate context. This runs contrary to influential views in philosophy of language that understand contextual backgrounding as implicit, unconscious and sub-intentional \citep{searle1980background, searle1983intentionality}, say as a kind of ability or \textit{knowledge-how} that allows interlocutors to interpret utterances appropriately and without explicit instructions for how to do so. For example, when a speaker utters \textit{Cut the cake}, hearers instinctively know to use a knife and not a lawnmower, even though this is left implicit. This is because of their background practical abilities, tendencies, and dispositions which generate the appropriate interpretation on demand. Secondly, in human conversation, priming is iterative and responsive: interlocutors adjust their framing based on real-time feedback from each other. By contrast, priming in the case of LLM prompting is more like a unilateral act of \emph{context imposition}, akin to a theatre director setting the stage, lighting, script, delivery, and success conditions for the performance in advance, while the actor (i.e., the LLM) must perform according to these specifications.

These asymmetries between human--human and human-LLM conversational priming reveal fundamental limitations in how context is scaffolded. While common ground in human-to-human conversation evolves dynamically, LLMs treat prompts as static contextual frames. Subsequent turns are interpreted strictly within this frame, even if the user's goals or other aspects of context shift mid-conversation. For example, a user who begins a conversation about climate policy but later pivots to historical precedents risks derailing the exchange unless they can somehow explicitly re-scaffold the context. Furthermore, users often overload prompts with contextual cues, leading to internal dilemmas or contradictions for the LLM in how to resolve the discourse topic or the goal of the conversation. Consider, for instance, the prompt:

\begin{quote}
\textit{Write a formal email to my boss apologizing for missing the deadline, but keep it casual---we're close friends.}
\end{quote}
Here the LLM is required to reconcile conflicting registers (i.e., formal/casual) and conflicting relational information (i.e., hierarchical/personal), often favoring statistically dominant or trained patterns (formality) rather than reasoning or possessing competence in how to resolve the tension. The epistemic opacity of which contextual cues the LLM has prioritized or ignored is also an issue. For example, a prompt like \textit{Explain quantum physics like I'm a five year old} may yield oversimplified analogies from the LLM, since it cannot verify whether the user actually possesses a five-year-old's knowledge base. This opacity stems from the LLM's lack of adequate \emph{pragmatic anchoring}---the ability to ground interpretations in shared perceptual or social context.

It could be said that something similar to context collapse also happens in prompt-driven human-LLM conversation, even though human--LLM conversation is not plausibly a form of mass communication like broadcast media. Context collapse is apparent in human--LLM conversation especially when observing how failures in scaffolding and in interpretation compound misunderstandings and compromise successful communication. For instance, in cases of scaffolding failure (whereby prompts inadequately specify context), models rely on training-data priors, blending incompatible norms. For example, a user asking for advice on a `work conflict' without specifying their industry might receive generic responses that conflate corporate, academic, and gig-economy dynamics. This mirrors Meyrowitz's \cite{meyrowitz1985no} observations about broadcast media flattening distinct audiences, except that here, the collapse arises from the model's inability to infer situational or domain specific boundaries from context, and the user's mistaken expectation that models possess human-like pragmatic capacities to do so. In a similar vein, LLMs can fail to recognize the user's conversational contributions and often misinterpret the illocutionary force of prompts. For example, a user's rhetorical question (e.g., \textit{Why would anyone believe the earth is flat?}) might be misread as a sincere request for an explanation of flat-earth theories. 
Such misalignments reflect the model's lack of metacommunicative competence: humans have the ability to reflexively negotiate meaning through shared context, tone, gesture, or clarifications like \emph{I'm being sarcastic}.

Context collapse in relation to prompt-driven priming, we surmise, stems from the model's lack of pragmatic and contextual anchoring, revealing LLMs' shortcomings when measured against the CONTEXT-ALIGN framework. Human conversation relies on shared perceptual environments, socio-cultural cues, real-time feedback, metacommunicative competence, and background to enable context-sensitive understanding and maintain the integrity of communication and context. LLMs, by contrast, simulate conversational context with surface-level features (like text- and pattern-recognition) but lack the analogue of mechanisms that underpin human communication, like common ground updating and pragmatic anchoring. To mitigate these issues and better meet the criteria outlined in the CONTEXT-ALIGN framework, LLM architectures must move beyond treating prompts as static contextual containers. This requires reimagining LLM conversational agents as collaborative partners in context-building rather than passive prompt executors. 

One mitigation strategy would be to allow conversational agents and users to engage in more explicit context negotiation---for example, by allowing users to query and adjust the model's perceived context (e.g., \textit{What do you think my goal is here?}). 
This would address scaffolding failures and overattribution of pragmatic understanding to LLMs. 
It would also put LLMs on a better footing with respect to criteria CA10 and CA2 of the CONTEXT-ALIGN framework by encouraging transparency in how the LLM is understanding and handling context. This would, in turn, ensure LLMs and users share more relevant assumptions and achieve better management of the common ground. Another mitigation strategy would be to increase LLM meta-pragmatic awareness by training them to flag ambiguous or contradictory cues in prompts (e.g., \textit{You mentioned both formal and casual tones---how should I prioritize these?}). 
This would help to address issues such as inconsistent prompts, overloading of the prompt, fragmentation of the conversation topic, and misrecognition of the conversational contribution. 
It would also put LLMs in a better position to meet the CONTEXT-ALIGN criteria CA5 and CA6. 
Finally, LLMs might be implemented to include more context verification (e.g., \textit{Just to confirm, you want examples relevant to Southeast Asia?}). 
This would allow the LLM to better integrate user corrections into context as it evolves, in line with CA4; and better track shifts in discourse referents, topics, and goals, in line with CA1 and CA3. 
Without such mechanisms, prompting will remain a brittle proxy for shared context---one that perpetuates undue burdens on users to exhaustively pre-empt misunderstanding and ensure contextual integrity. 
The design and implementation of LLMs should aim to replace the \emph{prompting as a substitute for context} paradigm by supporting architectures, operational constraints, and affordances that address the criteria in the CONTEXT-ALIGN framework.

\subsection{Behavioral Alignment and the Static Communicative Identity of LLMs} \label{subsec:5.2}

The behavioral alignment of LLMs, such as training them to be more disposed to behaving in accordance with predefined norms prescribed by the HHH framework, ensures they operate within certain ethical and pragmatic guardrails. However, this process is a poor approximation of more general pragmatic norms of human communication. Moreover, it imposes a static communicative identity on models that resists the situated normativity of (Gricean) pragmatics and the contextual tailoring of speaker identity fundamental to human communication. While humans dynamically adapt their conversational norms, register, and self-presentation to suit shifting contexts (e.g., switching from casual banter with a friend to formal discourse with a dean), LLMs are constrained by relatively fixed values embedded during system-prompting, and pre- and post-training (e.g., fine-tuning, reinforcement learning, character training). This rigidity creates a form of \emph{pragmatic dissonance}, where models mechanically enforce global norms (e.g., HHH) even when local contexts demand tailoring and flexibility.

Modern LLMs are shaped by different kinds of alignment strategies, each with distinct strengths and limitations. System prompts provide hard-coded directives (e.g., \textit{You are a friendly assistant\ldots}) that ensure consistency in tone and values, but also lock models into predefined personas that can stifle adaptability to local context. For instance, a model pre-prompted to `avoid controversial statements' might refuse to analyze partisan debates, even when contextually appropriate. Similarly, when models are fine-tuned, they are trained on domain-specific datasets (e.g., medical dialogues, customer service transcripts, legal jargon) to customize their behavior for particular domains or tasks. Fine-tuning increases pragmatic proficiency and alignment for these niche contexts, however there are risks of overfitting to narrow datasets, and reducing flexibility in general or inter-domain conversations. For example, a model fine-tuned on a data-set of academic writing might struggle with casual dialogue, rigidly adhering to formal registers, even though there may be contexts where such flexibility is required for alignment or would be desirable. With Reinforcement Learning (RL), models are optimized to prioritize outputs that human or AI raters deem friendly, helpful, honest, or harmless. RL assures broad ethical alignment with these virtues so that the LLM will avoid harmful outputs for example. However, this method and others like it prioritizes broad, universal norms over context-specific ones, often at the cost of pragmatic nuance or the specification of local context. For example, Claude's adherence to HHH norms might lead it to reject simplifying complex topics for children (\textit{I must provide accurate information}), favoring honesty over accessibility where this would otherwise be contextually required or desirable. 

Grice has taught us that human communication relies on situated normativity, the ability to flexibly apply, suspend, violate, or negotiate conversational rules and principles based on context. This reflexivity is relevant to CA6 and CA7. In the case of CA6, familiar from Grice, speakers are able to infer what is communicated (implicatures, explicatures, and figurative speech) by inferring which maxims are violated by the speaker in context. In the case of CA7, where the balance of ethical and pragmatic norms are at issue, LLMs need to be capable of reasoning to resolve normative conflicts in context. For example, when a doctor withholds a terminal diagnosis from a patient, they violate the maxim of quantity to uphold the norm of compassion demonstrating capacity to successfully navigate the combination of ethical and pragmatic norms that apply in context and resolve an appropriate manner of adherence to them in that context. 

Such situated normativity is also crucial to issues of speaker identity and self-presentation familiar from the discussion of context collapse and relating to CA8. 
LLMs lack the situated normativity and flexibility in self-presentation crucial to human communication. Their ``self-presentation'' is a corporate artifact of the way they are trained using alignment techniques (e.g., system prompting, character training), not a result of social and pragmatic negotiation. To illustrate, consider two contrasting scenarios: A professor jokes self-deprecatingly with colleagues at a conference dinner but adopts a formal tone during their keynote. 
Here, the professor engages in \emph{dynamic facework} with the ability to shape and adapt his character to suit different normative contexts. 
An LLM, however, fine-tuned for broad values such as neutrality, might refuse to engage in humor, even when the user signals a desire for levity (\textit{I'm sorry, I can't generate jokes}). 
LLMs, lacking face, default to a generic corporate persona, violating CA8 by applying one-size-fits-all norms (e.g., HHH) to all interactions. 
This is similar to context collapse on social media whereby speakers must present a contextually generic persona to appeal to or be understood by multiple audiences online. 

The static identity of LLMs exacerbates context collapse in ways that directly contravene the CONTEXT-ALIGN framework. 
They violate CA8, by applying the same alignment norms universally, flattening context-specific expectations. 
For example, a user seeking advice on office politics might receive sanitized responses (\textit{Focus on teamwork!}) that ignores the tacit and nuanced norms of workplace diplomacy because the LLM fails to recognize situational nuances or adapt to their audience. 
Furthermore, users cannot reshape LLM values or persuade them to follow different norms through dialogue. 
For instance, persuading an LLM to adopt a sarcastic tone is not possible without jailbreaking---a stark contrast to human conversations, where persuasion is rife and identities are co-constructed (e.g., bonding over shared humor). 
This undermines CA2 since it limits the ability to co-construct common ground, and CA11 since LLMs cannot evolve their characteristics across interactions to form enduring relationships with the user \citep{manzini2024code, kirk2025why}.\footnote{The development of AI companions is a relevant use-case.}  

Addressing these limitations requires rethinking alignment paradigms, LLM architectures, and affordances. One way of doing so would be to implement context-aware norm weighting to allow models to adjust norm prioritization as a function of context. 
Another strategy would be to allow some degree of user-driven persona customization allowing users to influence the models' values or alignment parameters as a function of context (e.g., \textit{Be 20\% more honest}). 
Finally, though AI safety experts might balk, it may be worthwhile to allow models to negotiate alignment, say, by allowing them to propose norm exceptions and adjust accordingly (e.g., \textit{This seems like a casual chat. Can I relax formality?}). 
Without innovations like these, LLMs will remain conversationally impoverished---ethically aligned yet pragmatically alien communicators---consistently out of sync with the fluidity of human communicative context.

\subsection{Summary}

To summarize, we claim prompting is a distinct kind of speech act that functions as a novel, yet problematic, substitute for the dynamic, communally-constructed context found in human--human conversation. 
We argue that this kind of static framing imposes challenges on achieving a shared understanding and conversational alignment between users and LLMs. 
We also argue that behavioral alignment strategies contribute to a rigid, pre-defined communicative identity for LLMs. 
This contrasts with the more flexible, context-sensitive forms of self-presentation found in human--human communication.

\section{Philosophical and Practical Implications} \label{sec:6}

In this paper, we have explored conversational alignment through the lens of contemporary theories of context, communication, and pragmatics, and identified central limitations within current LLM architectures, operational constraints, and alignment strategies. A fundamental philosophical implication of this analysis is that, despite their impressive successes in text-generation, LLMs do not yet fully qualify as human-like communicative agents in the more robust, pragmatic sense required by philosophical theories of meaning, context, and conversation. Indeed, even if such systems are ethically aligned and well fine-tuned, their pragmatic and conversational shortcomings are rife and fall well short of satisfying the desiderata outlined in the CONTEXT-ALIGN framework.

Having critically assessed some fundamental aspects of the conversational capacities of contemporary LLMs, we are now in a better position to address the key questions posed in the introduction and to summarize the philosophical significance of the foregoing. More specifically, we asked: What does it mean for a conversational agent to be aligned with human communicative practices? Do LLMs follow the norms, structures, and pragmatic expectations that govern human conversation? How do LLMs differ from human speakers in their conversational abilities? Are current LLM architectures, operational constraints, and affordances fundamentally limited in achieving full conversational alignment or can they be improved to better approximate human dialogue? We now answer these questions in turn. 

First, conversational alignment requires more than just the ability for text-generation; it requires a robust sensitivity to contextual parameters, the capacity to dynamically and cooperatively update the common ground, adherence to pragmatic norms, the effective management of discourse structure, and the flexibility to adapt one's communicative identity according to situation-specific demands. Conversationally aligned agents should function not merely as passive responders, but as active, cooperative participants in the joint construction of meaning, understanding, and communicative goals.

Second, our analysis indicates that current LLMs exhibit significant deficiencies when evaluated against the conversational alignment criteria outlined in our CONTEXT-ALIGN framework. While they can superficially approximate human conversation through learned statistical patterns, they may struggle to reliably track and update the common ground, to maintain discourse coherence by tracking QUDs, and they may often misinterpret the pragmatic force of utterances. Furthermore, the conversational norms and communicative identities imposed through static alignment strategies, such as RLHF and system prompts, lack the situated adaptability that characterizes human pragmatic competence. That is, rather than dynamically tailoring their conversational behaviors to the needs of the context, LLMs typically default to generic, inflexible personas that reflect overarching corporate or institutional values rather than context-specific requirements. Thus, although LLMs follow certain basic linguistic and ethical norms of human conversation, their adherence to deeper pragmatic norms, structures, and expectations remains limited and problematic.

Third, LLMs differ fundamentally from human interlocutors in several key dimensions of conversational competence. Humans possess genuine intentionality and meta-communicative awareness, enabling them to dynamically negotiate meaning and clarify misunderstandings through real-time feedback and conversational repair. In contrast, LLMs lack intentionality and meta-communicative competence, relying instead on static prompting and statistical inference from training data. For example, humans naturally maintain and update conversational context, common ground, and conversational scoreboards through situated pragmatic reasoning, whereas the architectural constraints of LLMs can result in significant challenges posed by the competing demands of context window overflow and context collapse, resulting in the loss of coherence and diachronic consistency over extended interactions. Furthermore, humans are capable of navigating shifts in communicative identity, norms, and registers across varying contexts, dynamically aligning their self-presentation and pragmatic norms to meet the situational demands. Conversely, LLMs that are constrained by static alignment protocols (such as HHH), exhibit rigid, context-insensitive communicative identities incapable of flexible adaptation. Thus, despite superficial linguistic fluency, LLMs differ from human speakers in their lack of genuine pragmatic reflexivity, meta-communicative negotiation, structured memory, and dynamic conversational identity.

Finally, our analysis suggests that current LLM architectures, constraints, and affordances may impose fundamental limitations on achieving full conversational alignment. The challenges of context window overflow and context collapse reveal inherent tensions between retaining sufficient context for coherence and avoiding pragmatic confusion. Similarly, static prompting as a substitute for context, combined with rigid alignment strategies, results in pragmatic dissonance and communicative identity inflexibility. We are inclined to think that these limitations are insurmountable in principle, although they might not be. In any case, they highlight the need for substantial conceptual and architectural innovation to move beyond current constraints.

There are good reasons to put in the effort to secure these conceptual and architectural innovations. We will end by discussing two significant practical implications of failing to meet the criteria of the CONTEXT-ALIGN framework. The first is that systematic failures in conversational alignment will inevitably have a disruptive effect on fundamental aspects of human communication, many of which are unpredictable and potentially negative in ways broadly analogous to the impacts that social media has had on communication (an example we have already discussed in detail with context collapse). The second is that systematic failure to meet the CONTEXT-ALIGN framework presents a significant safety issue. Misalignments in pragmatic understanding, conversational coherence, and context-tracking can lead to misunderstanding, misinformation, and erosion of user trust, as well as potentially giving rise to harmful consequences in high-stakes domains like healthcare, emergency response, education, or legal advice.

LLMs are not the first technology to transform fundamental aspects of communication. Email replaced much formality with asynchronous brevity, whereas social media turned discourse into a performative spectacle. Yet LLMs introduce a novel disruption: synthetic interlocutors that mimic human conversation while operating with wholly alien cognition, alien semantic contents, and alien principles. Our discussion reveals three important shifts that are beginning to emerge in human communication as a result of engaging extensively with conversational AI. As discussed, there is an emerging shift from collaborative, co-constructed conversation toward unilateral, prompt-driven conversation. There is also a shift in the norms governing communicative identity and self-presentation. Finally, we observe a shift toward epistemic opacity and diminished trust in conversational interaction. For while human interlocutors typically have some insight into each other's cognitive and communicative states---say through explicit requests for clarification, meta-communicative signaling, and pragmatic inference---LLMs, by contrast, provide little transparency into their internal reasoning, pragmatic assumptions, or contextual interpretations. 

Ultimately, like in the case of social media, addressing these shifts and their implications requires urgent philosophical reflection and technical innovation. The challenge is not simply to improve LLMs' linguistic fluency, intelligibility, and ethical alignment, but rather to ensure that their integration into our communicative lives respects---and ideally enhances---the pragmatic richness, flexibility, and cooperative spirit that define human conversation.

\section*{Acknowledgements}

We would like to thank Michael Barnes, Melissa Fusco, Karen Lewis, Eliot Michaelson, Eleonore Neufeld, Kate Scott, and the other participants of the \textit{Social Media and Communication} workshop held at Columbia University for helpful comments and discussion on this paper. We would also like to thank Adam Bradley, Iason Gabriel, Geoff Keeling, Seth Lazar, Fei Song, Kate Vredenburgh, and audience members at \textit{The Philosophy of AI: Themes from Iason Gabriel} workshop held at the AI \& Humanity Lab, University of Hong Kong.

\renewcommand{\bibsection}{\section*{References}}
\setlength{\bibsep}{0.0pt}
\bibliographystyle{plainnat}
\bibliography{main}

\end{document}